\documentstyle[psfig]{mn}

\footnotesize
\newdimen\digitwidth    
\setbox0=\hbox{\rm0}
\digitwidth=\wd0
\catcode`!=\active
\def!{\kern\digitwidth}
\normalsize

\def\dm{\hbox{pc cm$^{-3}$}}

\newcommand\nuddd{\ifmmode\stackrel{\bf \,...}{\textstyle\nu}\else$\stackrel{\,...}{\textstyle \nu}$\fi} 
\def\lsim{~\rlap{$<$}{\lower 1.0ex\hbox{$\sim$}}}
 
\title{Long-term Scintillation Observations of Five Pulsars at 1540~MHz}

\author[Wang, Manchester et al.]{N. Wang,$^{1,2,3}$\thanks{Email:
na.wang@ms.xjb.ac.cn} ~~R. N. Manchester,$^{3}$ 
~~S. Johnston,$^2$  ~~ B. Rickett,$^{4}$ ~~J. Zhang,$^1$
\newauthor A. Yusup,$^{1}$ ~~M. Chen$^{1}$\\
$^{1}$ National Astronomical Observatories, CAS, 40-5 South Beijing Road, Urumqi, 830011, China\\
$^{2}$ School of Physics, University of Sydney, NSW 2006, Australia \\ 
$^{3}$ Australia Telescope National Facility, CSIRO, PO Box 76, Epping, NSW 1710, Australia\\
$^{4}$ Department of Electrical and Computer Engineering, University of California, San Diego, USA\\}

\begin{document}
\maketitle
\pagestyle{plain}

\begin{abstract}
From 2001 January to 2002 June, we monitored PSRs B0329+54, B0823+26,
B1929+10, B2020+28 and B2021+51 using the Nanshan 25-m radio telescope
of Urumqi Observatory to study their diffractive interstellar
scintillation (DISS). The average interval between observations was
about 9 days and the observation duration ranged between 2 and 6 hours
depending on the pulsar. Wide variations in the DISS parameters were
observed over the 18-month data span. Despite this, the average
scintillation velocities are in excellent agreement with the proper
motion velocities. The average two-dimensional autocorrelation
function for PSR B0329+54 is well described by a thin-screen
Kolmogorov model, at least along the time and frequency axes.
Observed modulation indices for the DISS time scale and bandwidth and
the pulsar flux density are greater than values predicted for a
Kolmogorov spectrum of electron density fluctuations. Correlated
variations over times long compared to the nominal refractive
scintillation time are observed, suggesting that larger-scale density
fluctuations are important. For these pulsars, the scintillation
bandwidth as a function of frequency has a power-law index ($\sim3.6$)
much less than expected for Kolmogorov turbulence ($\sim4.4$).
Sloping fringes are commonly observed in the dynamic spectra,
especially for PSR B0329+54. The detected range of fringe slopes are
limited by our observing resolution. Our observations are sensitive to
larger-scale fringes and hence smaller refractive angles,
corresponding to the central part of the scattering disk. 

\end{abstract}


\section{Introduction}
Diffractive interstellar scintillation (DISS) occurs when radiation
from a pulsar has phase fluctuations imposed on it due to small scale
variations in the interstellar electron density. The scattered rays
 interfere constructively and destructively, resulting in a
modulation of the pulse intensity as a function of frequency and
position on the observer plane. Pulsars are good probes of DISS
because of their very small angular size. The relative motion of the
pulsar, scattering material and the observer transforms the spatial
variation into a time variation of the signal intensity. The
modulation is characterized by a timescale ($\Delta t_{\rm d}$) of minutes
to hours and a decorrelation frequency scale ($\Delta \nu_{\rm d}$)
over a wide frequency range from kHz to MHz, with distant pulsars
scintillating faster in both frequency and time
\cite{cpl86,grl94,brg99}. Longer-term fluctuations in pulsar flux
densities result from a focussing and defocussing of the ray bundle due to
larger-scale inhomogeneities in the interstellar electron
density. This phenomenon, known as refractive interstellar
scintillation (RISS), is broad-band and shows less modulation in distant
pulsars.

By analogy with neutral gas turbulence theory, the density fluctuations in
the ionized interstellar medium (ISM) can be described by a power-law
spectrum.  The three-dimensional spatial power-law spectrum is of the
form:
\begin{equation} 
P_{3n}=C^2_n\,q^{-\beta}  
\label{eq:kolmogorov}
\end{equation}
where $C^2_{\rm n}$ is a measurement of the mean turbulence of
electron density along the line of sight and $q=2\pi/s$ is the
wavenumber associated with the spatial scale of turbulence $s$. The
spatial scale $s$ is in the range of $s_{\rm inn}\ll s\ll s_{\rm
out}$, where $s_{\rm inn}$ and $s_{\rm out}$ correspond to the inner
and outer scales of the  density fluctuations.  The index $\beta$ is
thought to lie in the range $3<\beta<5$. For propagation of turbulent
energy from large scales to small scales, the well-known Kolmogorov
theory of turbulence gives a value of $\beta=11/3$. Observations show
that this generally applies in the interstellar medium between scales
of $s_{out}\sim10^{17}-10^{19}$~cm and $s_{\rm inn}\sim
10^{8}-10^{10}$~cm \cite{sg90,ks92,ars95,bgr99}. 

There are discrepancies in the observed scintillation spectra compared
to the predictions of Kolmogorov turbulence \cite{grc93}. For example,
modulation depths and time scales often do not scale according to the
Kolmogorov model. Various modifications to the Kolmogorov model have
been proposed, including the steep spectrum model ($\beta\geq4$)
\cite{bn85,gn85} and the inner scale model are notable. The steep
spectrum model does not require a turbulent cascade; it could result
from the superposition of noninteracting discrete structures in
space. The inner scale model has a cutoff at the scale at which the
turbulent energy dissipates \cite{cfrc87}.

The variation in observed pulsar flux density as a function of time
and frequency is known as the dynamic spectrum and the interference
maxima in the dynamic spectrum are called scintles.  Decorrelation
time scale and bandwidth are parameters which quantify 
the scintle size. A simplified model is often used in which a thin scattering disk
is assumed to lie midway between the pulsar and observer. In general
this is in good agreement with observations.  For a pulsar at distance
$D$ from the Earth, the decorrelation time scale for DISS has the form
of
\begin{eqnarray}
\Delta t_{\rm d}\propto \nu^{2/(\beta-2)}\,D^{-1/(\beta-2)}\,V_{\rm s}^{-1} & (\beta<4),
\label{eq:td1} 
\end{eqnarray}
\begin{eqnarray}
\Delta t_{\rm d}\propto \nu^{(\beta-2)/(6-\beta)}\,D^{-(\beta-3)/(6-\beta)}\,V_{\rm s}^{-1} & (\beta>4),
\label{eq:td11} 
\end{eqnarray}
where $\nu$ is the radio frequency \cite{ric77,gn85}.  In particular, for a Kolmogorov spectrum, 
\begin{equation}
\Delta t_{\rm d} \propto \nu^{6/5}D^{-3/5}\,V_{\rm s}^{-1}.
\label{eq:td2} 
\end{equation}
The decorrelation bandwidth for DISS is
\begin{eqnarray}
\Delta\nu_{\rm d}\propto \nu^{2\beta/(\beta-2)}\,D^{-\beta/(\beta-2)} & (\beta<4),
\label{eq:dnu1} 
\end{eqnarray}
\begin{eqnarray}
\Delta\nu_{\rm d}\propto \nu^{8/(6-\beta)}\,D^{-\beta/(6-\beta)} & (\beta>4)
\label{eq:dnu11} 
\end{eqnarray}
\cite{ric77,gn85}.  In the Kolmogorov case,
\begin{equation}
\Delta\nu_{\rm d} \propto \nu^{22/5}\,D^{-11/5}.
\label{eq:dnu2}  
\end{equation}
Equations~\ref{eq:td1} to \ref{eq:dnu2} indicate that fluctuations are more
rapid and have narrower bandwidths for more distant pulsars and at
lower frequencies.

In the Kolmogorov case, the scaling factor $C_{\rm n}^2$ is given by
\begin{equation}
C_{\rm n}^2\approx0.002\,\nu^{11/3}D^{-11/6}\Delta \nu_{\rm d}^{-5/6},
\label{eq:cn2}
\end{equation}
where $\nu$ is in GHz, $D$ is in kpc, $\Delta\nu_{\rm d}$ is in MHz
and $C_{\rm n}^2$ is in units of m$^{-20/3}$ \cite{cor86}.

Quasi-periodic fringes resulting from interference between multiple images of the pulsar
by the scattering screen, are often observed in dynamic spectra
\cite{hwg85,wc87,brg99}, with broad and narrow fringes indicating small
and large separations of the ray paths (or images) respectively. The
fringes are generally sloped on the dynamic spectra, i.e., periodic in
both frequency and time and crossed fringes (i.e., slopes of both
signs) are common. Such sloped fringes result from motion of the
fringe pattern across the observer plane due to relative motion of the
pulsar, Earth and the scattering screen. In the case of an equivalent
thin screen located approximately midway between the pulsar and the
observer, the maximum fringe slope is given by
\begin{equation}
\frac{dt}{d\nu} = \frac{D\,\theta_{\rm r}}{V_{\rm s}\nu}
\label{eq:fringe}
\end{equation}
where $\theta_{\rm r}$ is the refraction angle and $V_{\rm s}$ is the
velocity of the fringe pattern across the observer plane, normally
 dominated by the pulsar velocity \cite{hew80,bgr99}. The
refractive angle $\theta_{\rm r}$ is approximately proportional to
$\nu^{-2}$ for a given gradient in refractive index. Therefore, for
large-scale gradients, the fringe slope $dt/d\nu$ might be expected to
vary approximately as $\nu^{-3}$. For different projections of the
velocity on the line between the images, the slope varies in magnitude
and sign, resulting in crossed fringes of varying slope. Earlier
papers discussed Equation~\ref{eq:fringe} in terms of the frequency
dependence of the refraction angle (e.g. Gupta, Rickett \& Lyne,
1994\nocite{grl94}), but this equation results directly from
consideration of the geometric delays. For interference between a
strong central image and other points within the midplaced scattering
disk, the fringe rates in frequency ($f_\nu$) and time ($f_{\rm t}$)
are related by
\begin{equation}
f_{\nu} = \frac{D}{2c\,\nu^2\, V_{\rm s}^2} f_{\rm t}^2,
\label{eq:arc}
\end{equation}
where $c$ is speed of light \cite{smc+01}, and $V_{\rm s}$ is
dominated by the pulsar proper motion. This shows that crossed fringe
patterns transform to parabolic arc structures in the `secondary
spectrum', that is, the two-dimensional Fourier transform of the
dynamic spectrum.  Such parabolic arcs and other more complex
structures have been observed in a number of pulsars \cite{hsb+03}.

In this paper, we describe observations of five pulsars (PSRs
B0329+54, B0823+26, B1929+10, B2020+28 and B2021+51).  In
Section~\ref{sec:obs} we introduce the observations and data
analysis. Section~\ref{sec:spc} presents the dynamic spectra and in
Section~\ref{sec:parfx} \& \ref{sec:modacf} we show the
auto-correlation functions and describe the time variations of the
scintillation parameters.  Secondary spectra for PSR B0329+54 are
presented in Section~\ref{sec:secsp}. In Section~\ref{sec:dis} we
discuss our results and compare them with previous observations and
the predictions of scintillation theories, and Section~\ref{sec:sum}
summarises the results.

\section{Observations and Data Analysis}\label{sec:obs}
The observations were made by using the Nanshan 25-m telescope,
operated by Urumqi Observatory, National Astronomical Observatories of
China.  The pulsars were regularly monitored from 2001 January to 2002
June with an average interval between observations of nine days. Our
observations were centred at 1540~MHz, a relatively high frequency at
which pulsar scintillation properties have not been well studied. Most
earlier observations (e.g., Cordes, Wesberg \& Boriakoff 1985; Gupta,
Rickett \& Lyne 1994; Bhat, Rao \& Gupta 1999) were less frequent and
centred on lower frequencies, 327, 408 and 610~MHz. The receiver was a
dual-polarisation room-temperature system with noise temperature of
95~K, equivalent to a system flux density of 1080~Jy.  Frequency
resolution was provided by a filterbank system consisting of 128
channels with channel bandwidths of 2.5~MHz giving a total bandwidth 
of 320~MHz for each polarisation.  This system was
described in more detail by Wang et al. (2001)\nocite{wmz+01}.  The
data were folded online and saved to disk every four minutes. Total
integration time was from three to six hours, depending on the pulsar
scintillation time scale $\Delta t_{\rm d}$. We made 6-hour
observations for the closer pulsars, PSRs B1919+10, B2020+28 and
B2021+51, while the observations were three hours for the more distant
pulsars PSRs B0329+54 and B0823+26. Between 38 and 65 observations
were made of each pulsar. We calibrated the pulsar flux density scale
using ten strong, relatively distant pulsars which all have
well-measured flux densities in the pulsar catalogue.

Parameters for the observed pulsars are listed in
Table~\ref{tb:eph}. Column~3 gives the pulsar dispersion measure and
column~4 gives the pulsar distances obtained from parallax
observations. Uncertainties in the last quoted digit are given in
parentheses. Column~5 gives the transverse velocities based on
measured proper motion and the parallax distance.  Columns~6 and 7
give the pulsar Galactic longitude and latitude respectively, and
pulsar distances from the Galactic plane are given in last column.

\begin{table*}
\begin{minipage}{150mm}
\caption{Parameters for the observed pulsars.}
\begin{tabular}{ccccccrr}
\hline    & \vspace{-3mm} \\
PSR J     & PSR B   & DM     
                                    
                           & Parallax dist &$V_{\rm pm}$      &  $l$     & $b$~~~ & $z$~~~ \\
          &         & ($\dm$) & (kpc)         & (km~s$^{-1}$)    & ($\deg$ )&($\deg$)&(kpc)\\
(1) & (2) & (3) & (4) & (5) & (6) & (7) & (8)  \\
\hline    & \vspace{-3mm} \\
0332+5434 & 0329+54 & 26.84  & 1.06(12)\footnote{Brisken et al. (2002)\nocite{bbgt02}} & ~98(11)$^a$  & 145.00 & --1.22 &--0.02\\
0826+2637 & 0823+26 & 19.45  & 0.38(8)\footnote{Gwinn et al. (1986)\nocite{gtwr86}}   & 194(41)\footnote{Lyne, Anderson \& Salter, (1982)\nocite{las82}}    & 196.96 &  31.74 &  0.20\\
1932+1059 & 1929+10 & ~3.18  & 0.33(1)$^a$                      & 163(5)$^a$    & ~47.38 & --3.88 &--0.02\\
2022+2854 & 2020+28 & 24.64  & ~2.7(9)$^a$                     & 307(103)$^a$  & ~68.86 & --4.67 &--0.22\\
2022+5154 & 2021+51 & 22.65  & ~2.0(3)$^a$                     & 119(18)$^a$   & ~87.86 &   8.38 &  0.29\\
\hline  & \vspace{-3mm} \\
\end{tabular}
\label{tb:eph}
\end{minipage}
\end{table*}

To get the dynamic spectrum, we first dedispersed and summed the data
for each observation to determine the pulse phase at the centre
frequency. Pulse intensities were then determined for each
sub-integration and each frequency channel by summing the data within
phase $\pm0.05$ of the predicted phase of the pulse peak and
subtracting a baseline offset.

Persistent narrow-band interference was found in frequency ranges
1465--1472~MHz, 1482--1490~MHz, 1522--1527~MHz, 1545--1557~MHz and
1620--1625~MHz; in total about 15\% of the frequency channels were
affected. We used a linear interpolation across affected channels to
remove the interference from the dynamic spectra.  Occasionally,
impulsive broad-band interference which lasted a few minutes was
observed; a similar interpolation was used to remove it.

\section{Dynamic Spectra}\label{sec:spc}
In this section we present dynamic spectra for PSRs B0329+54,
B0823+26, B1929+10, B2020+28 and B2020+51. Spectra selected to show
the range of observed features for each pulsar are given in
Fig.~\ref{fg:0329_spc} to Fig.~\ref{fg:2021_spc}. The observing start
time, pulsar name and MJD are given at the top of the each spectrum.

\subsection{PSR B0329+54}
\begin{figure*}
\begin{center} 
\mbox{\psfig{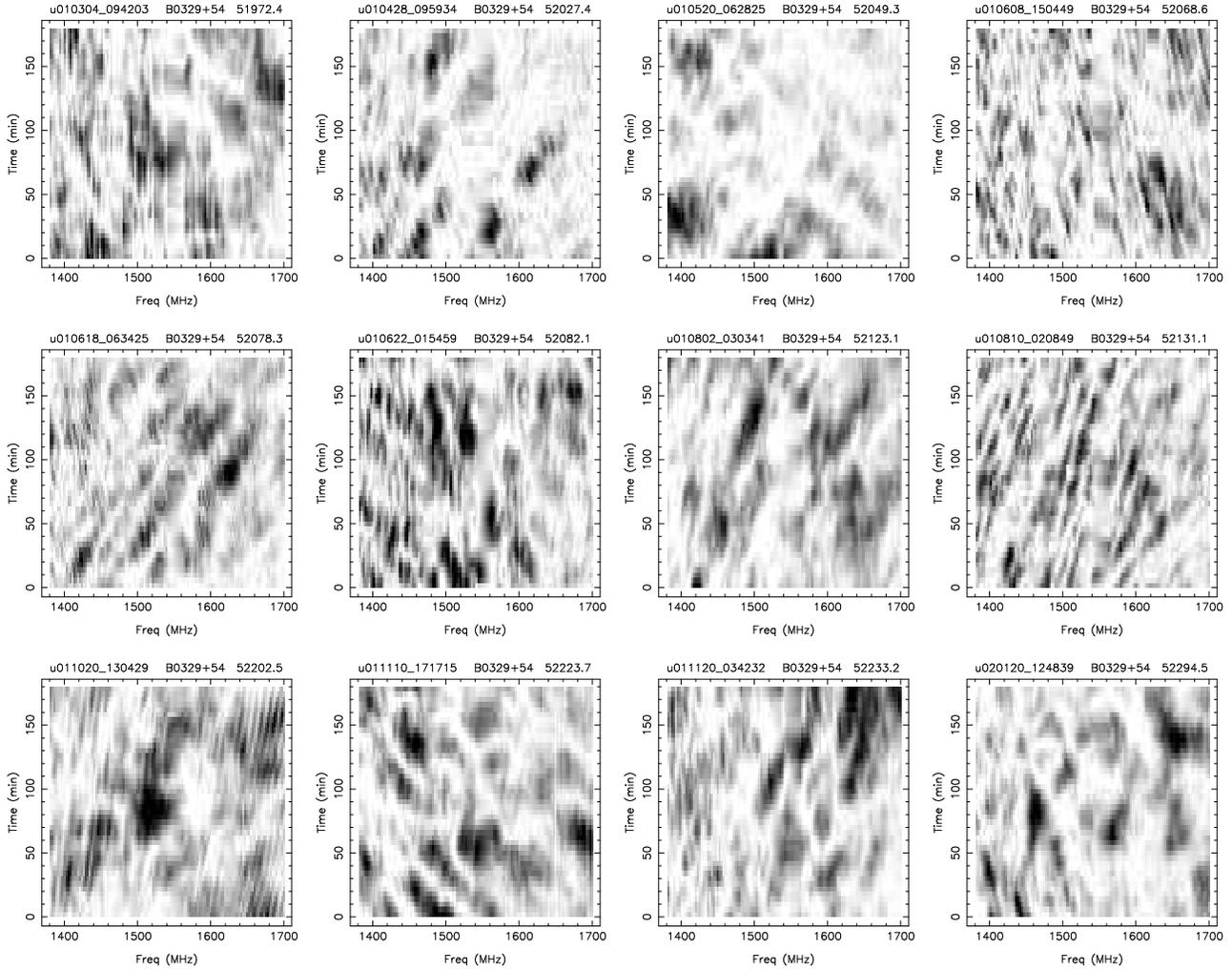}}
\caption{Dynamic spectra for PSR B0329+54 at center frequency 1540~MHz,
with frequency and time resolution of 2.5~MHz and 4~min
respectively. Intensity is represented by a linear greyscale between
1\% (white) and 85\% (black) of the maximum.}
\label{fg:0329_spc}
\end{center}
\end{figure*}

PSR  B0329+54 is the strongest northern pulsar with a distance 
of 1.06~kpc. At 1540~MHz, the pulsar scintillates with
typical time scales of 10 -- 30~min and decorrelation bandwidths
of 5 -- 15~MHz. We observed PSR B0329+54 at 64 epochs over the
18-month interval, with a typical observation time of three
hours. Fig.~\ref{fg:0329_spc} shows 12 of the dynamic spectra which
have been corrected for both the short broad-band interference and the
longer-term narrow-band interference.
  
The dynamic spectra shown in Fig.~\ref{fg:0329_spc} reveal very
different properties compared to the lower frequency observations
reported by Gupta, Rickett \& Lyne (1994) and Stinebring, Faison \&
McKinnon (1996) \nocite{grl94,sfm96} which were at 408~MHz and 610~MHz
respectively. There is a much wider variation in scintillation
time-scales and bandwidths from day to day in the 1540~MHz data and
many spectra show multiple fringes, which are usually sloping, for
example, the dynamic spectra obtained on MJDs 52068.6, 52078.3,
52082.1, 52131.1 and 52202.5.  Broad and narrow fringes 
are often seen at the same time, and these often have opposite
slopes. Examples are spectra for MJDs 52027.4, 52078.3 and
52082.1. The dynamic spectrum for MJD 51972.4 is interesting in that
it has narrow vertical fringes.

\subsection{PSR B0823+26}
\begin{figure*}
\begin{center} 
\mbox{\psfig{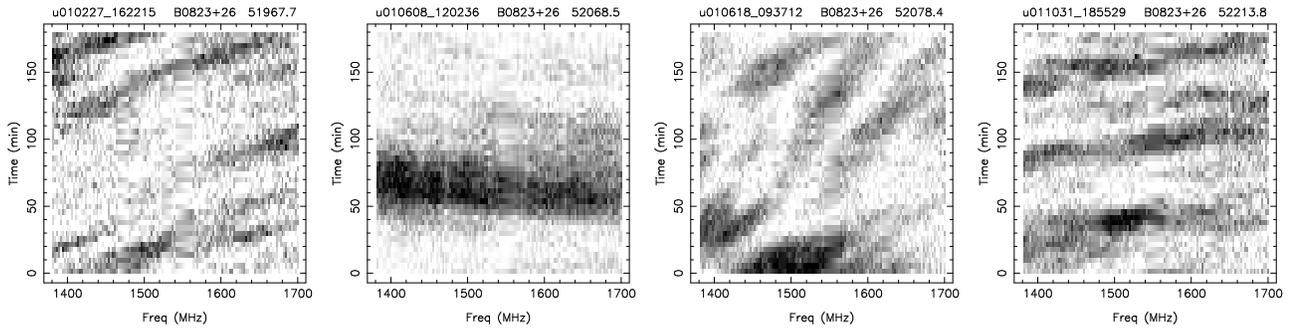}}
\caption{As in Fig.~\ref{fg:0329_spc}, dynamic spectra for PSR
B0823+26.}
\label{fg:0823_spc}
\end{center}
\end{figure*}

In total we obtained 48 dynamic spectra for PSR B0823+26, and
Fig.~\ref{fg:0823_spc} shows 4 of them. This pulsar has
 a distance of 0.38~kpc and a large transverse velocity of
194~km~s$^{-1}$. These parameters result in a wide decorrelation
bandwidth and a short decorrelation time scale.  As will be discussed
in Section~\ref{sec:dis}, the average scintillation scale of PSR
B0823+26 is about 13~min and the decorrelation bandwidth is about half
the receiver bandwidth. As seen with other pulsars the scintle
time-scale is shorter than the interval between scintles at a given
frequency.

Despite the frequency difference between our observations and those of
Gupta, Rickett \& Lyne (1994) at 408~MHz and Bhat, Rao \& Gupta (1999)
at 327~MHz, the dynamic spectra shown in Fig.~\ref{fg:0823_spc} are
similar with large fractional decorrelation bandwidths and slow
drifting. The character of the dynamic spectra often change
dramatically over short time intervals, for example, from a very broad
single scintle to multiple sloping scintles in observations on MJDs
52068.5 and 52078.4 respectively. Clear reversals in the direction of
drift slope are seen in observations made just one week apart, e.g.,
those on MJDs 52249.0 and 52256.0.

\subsection{PSR B1929+10}
\begin{figure*}
\begin{center} 
\mbox{\psfig{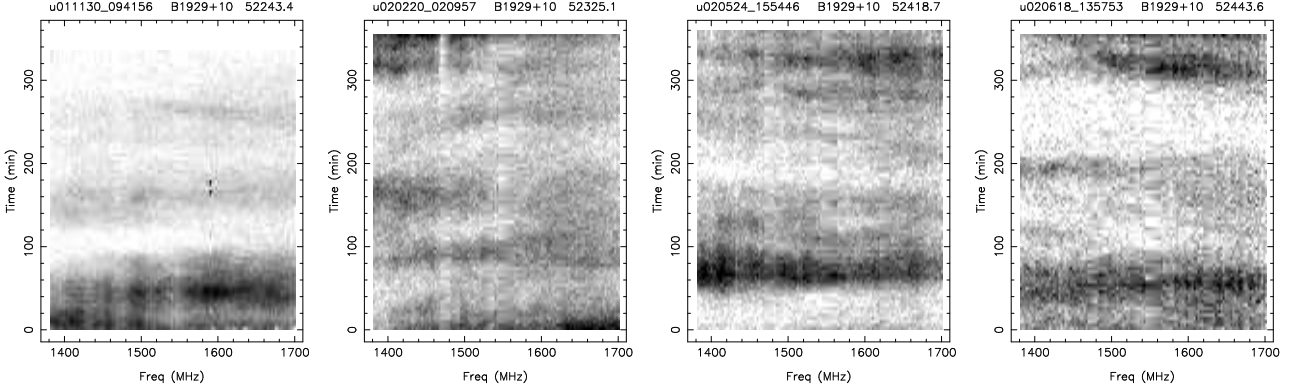}}
\caption{As in Fig.~\ref{fg:0329_spc}, dynamic spectra for PSR
B1929+10. Note the different time scale on the vertical axis compared
to Fig.~\ref{fg:0329_spc}.}
\label{fg:1929_spc}
\end{center}
\end{figure*}

We obtained dynamic spectra for PSR B1929+10 at 38 epochs. This
pulsar, with distance of 0.33~kpc,
scintillates slowly and its decorrelation bandwidth is obviously
comparable to or wider than the receiver bandwidth.  As shown in
Fig.~\ref{fg:1929_spc}, the six-hour observations typically were able
to cover only one or two scintles in the time domain. Features of the
spectra at 1540~MHz are very similar to those observed at lower
frequencies by Gupta, Rickett \& Lyne (1994). \nocite{grl94}

\subsection{PSR B2020+28}
\begin{figure*}
\begin{center} 
\mbox{\psfig{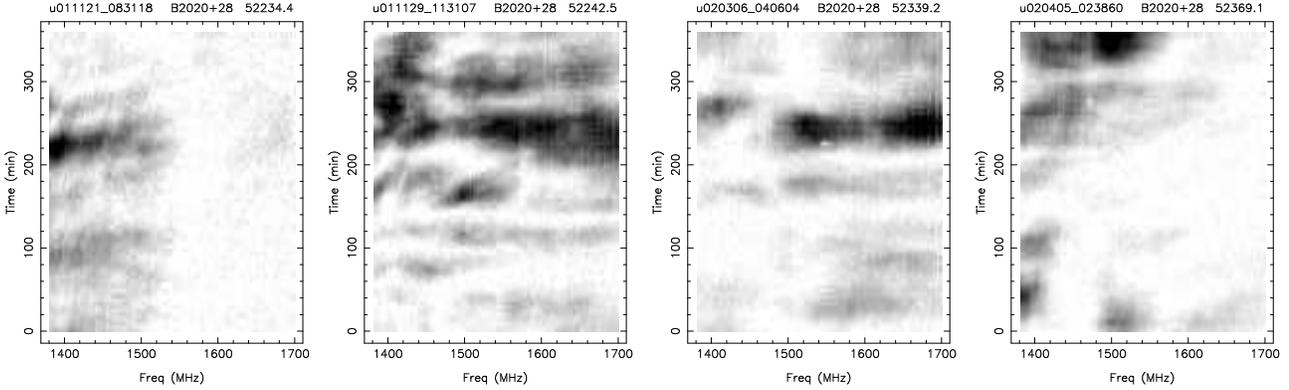}}
\caption{As in Fig.~\ref{fg:0329_spc}, dynamic spectra for PSR
B2020+28.}
\label{fg:2020_spc}
\end{center}
\end{figure*}

We observed 52 dynamic spectra for PSR B2020+28. At earlier stages
most of the observations were about 3 hours but later we commenced
6-hour observations in order to cover several scintles in each
observation. In Fig.~\ref{fg:2020_spc} we present 4 dynamic spectra.
Despite its DM of 24.6~$\dm$ and a distance of 2.7~kpc, PSR B2020+28
shows rather broad scintles in both time and frequency.

\begin{figure*}
\begin{center} 
\mbox{\psfig{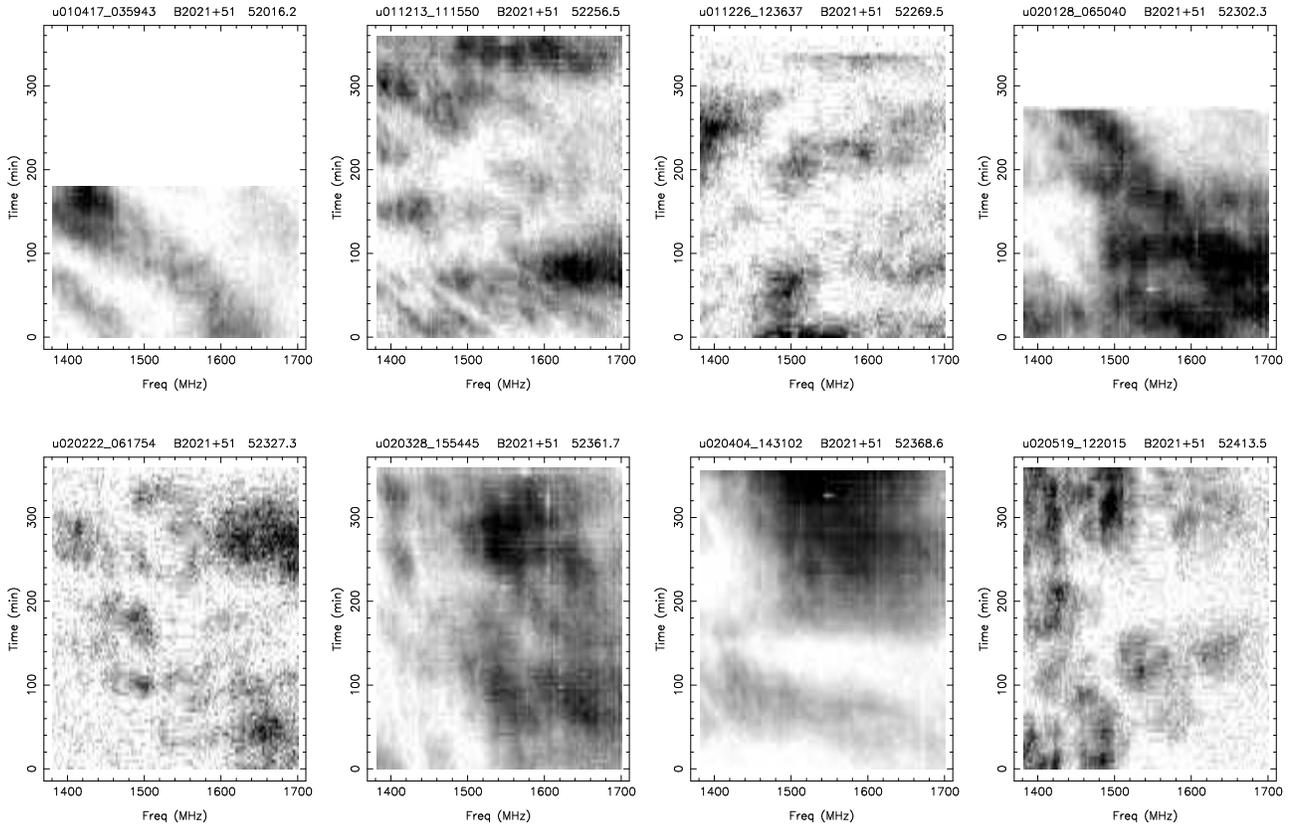}}
\caption{As in Fig.~\ref{fg:0329_spc}, dynamic spectra for PSR
B2021+51.}
\label{fg:2021_spc}
\end{center}
\end{figure*}

\subsection{PSR B2021+51}
As for PSR B2020+28, PSR B2021+51 initially was observed for three
hours each day; from mid-November 2001 observation times were
increased to six hours. Altogether we observed 65 dynamic spectra for
this pulsar, and 8 examples are shown in
Fig.~\ref{fg:2021_spc}. Clear drifting bands were observed on several
occasions, for example, the first observed dynamic spectrum at MJD
52016.2. This pulsar shows two types of dynamic spectrum. In some
cases the scintles are broad in both frequency and time (e.g., at MJD
52302.3), whereas in other cases the scintles are narrower (e.g., at
MJD 52269.5). There appears to be a correlation between signal
strength and scintillation type, with the pulsar generally being
weaker when the scintles are narrow.

\section{Scintillation parameters and flux densities }\label{sec:parfx}
\subsection{Auto-correlation functions}\label{sec:acf}
The parameters of DISS can be obtained by forming a two-dimensional
auto-correlation function (ACF) of the dynamic spectra $S(\nu,t)$,
i.e., the pulsar flux density at frequency $\nu$ and time $t$. The ACF
is defined by
\begin{eqnarray}
F(\Delta\nu, \Delta t) = \nonumber \\
\left[N(\Delta\nu,\Delta t)\right]^{-1}\sum_\nu \sum_t \Delta S(\nu,t) \Delta S(\nu+\Delta\nu,t+\Delta t),
\label{eq:acf} 
\end{eqnarray}
where $\Delta S(\nu,t) = S(\nu,t)-\bar{S}$, and $\bar{S}$ is the mean
pulsar flux density over each observation. $N(\Delta\nu,\Delta t)$ is the
number of correlated pixel pairs. The normalised ACF is then given by: 
\begin{equation} 
\rho(\Delta\nu,\Delta t) = F(\Delta\nu,\Delta t)/F(0,0). 
\label{eq:norm}
\end{equation} 

We note that in Cordes (1986) \nocite{cor86} the flux density was
normalised by subtracting the mean over each sub-integration to
minimise the effect of long-term variations, i.e., $\Delta S(\nu,t) =
S(\nu,t)-\bar{S}(t)$.  However this method can only be applied to
spectra that contain many scintles, since in this case the
sub-integration mean is close to the global mean apart from long-term
variations. Because of our higher observation frequency, the number of
scintles is often small and subtracting the mean from each
sub-integration would affect the diffractive parameters.

\begin{figure*}
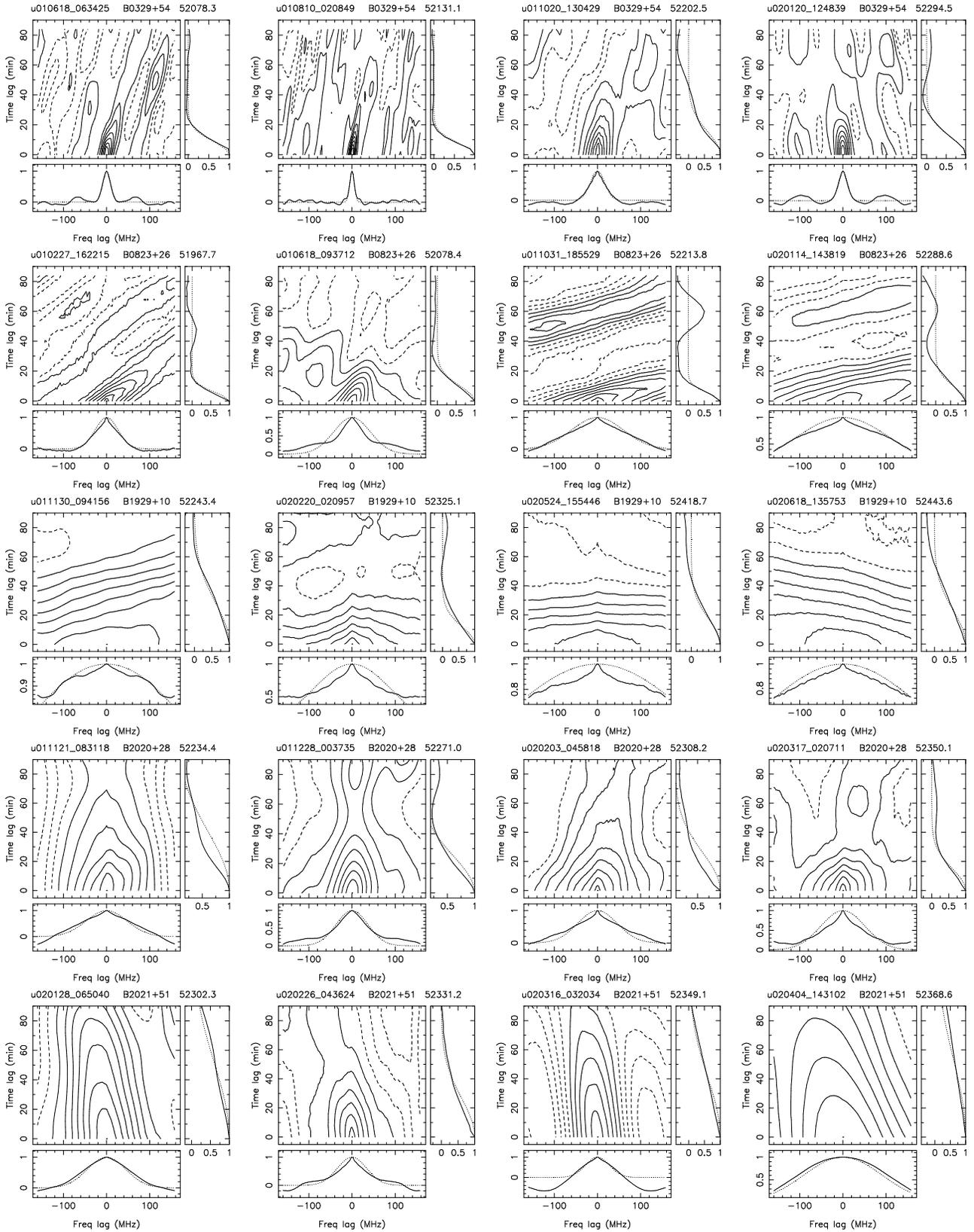

\begin{center} 
\begin{tabular}{l}
\mbox{\psfig{file=0329_acf.ps,width=170mm,angle=270}}  \\
\mbox{\psfig{file=0823_acf.ps,width=170mm,angle=270}}  \\
\mbox{\psfig{file=1929_acf.ps,width=170mm,angle=270}}  \\
\mbox{\psfig{file=2020_acf.ps,width=170mm,angle=270}}  \\
\mbox{\psfig{file=2021_acf.ps,width=170mm,angle=270}}  \\
\end{tabular}
\caption{Samples of normalised two-dimensional auto-correlation
functions of the dynamic spectra shown in Fig.~\ref{fg:0329_spc} to
Fig.~\ref{fg:2021_spc}. The contour interval is 0.1, with dashed lines
representing negative contour levels.  The solid line in the two smaller
plots at the bottom and right are the one-dimensional ACFs at zero lag
in time and frequency respectively. The dotted lines are zero-lag cuts
through the two-dimensional Gaussian fits to the observed data.}
\label{fg:acf}
\end{center}
\end{figure*}

Four examples of normalised two-dimensional ACFs and one-dimensional
cuts at zero lag are plotted in Fig.~\ref{fg:acf} for each of the five
pulsars. Following convention, the scintillation time scale $\Delta
t_{\rm d}$ is defined as the time lag at zero frequency lag at the
point where the ACF is 1/e of the maximum, and the decorrelation
frequency scale $\Delta\nu_{\rm d}$ is defined as the half-width at
half-maximum of the ACF along the frequency lag axis at zero time lag
\cite{cor86}.  The ACFs show spikes at zero time and frequency
lag. These spikes are due to noise in the dynamic spectra, and are
more significant for weaker pulsars. Slightly wider spikes resulting
from the interference excision process are often also present around
zero frequency lag. To remove these spikes we used the neighbouring
frequency-lag points to fit for a parabola across zero frequency
lag. The central points are interpolated from this fit.

\subsection{Scintillation parameters}\label{sec:para}
We estimated the diffractive scintillation parameters for each
observation by fitting a two-dimensional elliptical Gaussian
function \cite{brg99} to the smoothed ACF  using the non-linear
least-squares fitting routine {\sc LMM}. The Gaussian
function has the form:

\begin{equation}
C_g(\nu,t)=C_0\,{\rm exp}(C_1\,\nu^2+C_2\,\nu\,t+C_3\,t^2),
\label{eq:gauss}
\end{equation}
where $C_0$ is held fixed to unity as the ACF is normalised. The
scintillation parameters $\Delta t_{\rm d}$ and $\Delta\nu_{\rm d}$ are
then given by:

\begin{equation}
\Delta t_{\rm d}=\left(\frac{1}{C_3}\right)^{0.5},
\Delta\nu_{\rm d}=\left(\frac{{\rm ln}2}{C_1}\right)^{0.5}.
\label{eq:para}
\end{equation}

Fig.~\ref{fg:acf} shows that the Gaussian function represents the
observed ACFs well in most cases, although it may not be very reliable when
the decorrelation bandwidth is wider than the receiver bandwidth, as
in most observations of PSR B1929+10. Time variations of $\Delta
t_{\rm d}$ and $\Delta \nu_{\rm d}$ from all observations of the five
pulsars are shown in Fig.~\ref{fg:0329_var} to Fig.~\ref{fg:2021_var}
and average values are given in columns 2 and 3 of
Table~\ref{tb:para}.  One-sigma error estimates from the least-squares
fit are plotted on each point in the figure. Errors in the table are
one-sigma uncertainties in the mean values, assuming a normal
distribution. There is an additional statistical error related to the
finite number of scintles in the dynamic spectrum. Following Cordes,
Weisberg \& Boriakoff (1985),\nocite{cwb85} we take the fractional
uncertainties in $\Delta t_{\rm d}$ and $\Delta \nu_{\rm d}$ to be
given by $N^{-0.5}$ where $N = T_{\rm obs} \Delta\nu_{\rm
obs}/(4\,\Delta t_{\rm d} \Delta\nu_{\rm d})$ is the number of
observed scintles. Typical values of these additional uncertainties
are shown by error bars in the upper left corner of the $\Delta t_{\rm
d}$ and $\Delta \nu_{\rm d}$ plots in Fig.~\ref{fg:0329_var} to
\ref{fg:2021_var}.

Fitting for the two-dimensional Gaussian ellipse provides parameters
for the sloping features of the dynamic spectrum.  Following Bhat, Rao
\& Gupta (1999), we adopt $dt/d\nu$ as the drift rate; in terms of
the fitted parameters $dt/d\nu$ is given by:
\begin{equation}
\frac{dt}{d\nu}=-\left(\frac{C_2}{2\,C_3}\right).
\label{eq:slope}
\end{equation}
The slope visibility $r$ is given by:
\begin{equation}
r=C_2/\sqrt{4C_{1}\,C_{3}}.
\label{eq:vis}
\end{equation}
Derived values for the drift rate and absolute value of slope
visibility ($|r|$) are plotted in Fig.~\ref{fg:0329_var} to
Fig.~\ref{fg:2021_var} and average values are given in columns 5 and 6
of Table~\ref{tb:para}.

The scattering strength $u$ is defined to be the ratio of Fresnel
scale $s_{\rm F}$ to the DISS scale $s_{\rm d}$ \cite{ric90,gup95};
$u\gg 1$ corresponds to strong scattering and $u\ll 1$ to weak
scattering. In terms of observable quantities, $u$ is expressed as
\begin{equation}
u=\frac{s_{\rm F}}{s_{\rm d}}\sim\left(\frac{2\nu}{\Delta \nu_{\rm d}}\right)^{0.5}
\label{eq:u}
\end{equation}
 \cite{bgr99}, where $\nu$ is the observing frequency. Observed mean
values of $u$ are given in column 7 of Table~\ref{tb:para}. Scattering
is stronger for PSR B0329+54 compared to PSRs B0823+26, B2020+28, and
B2021+51, despite their similar DMs.

Based on the simple thin screen model with power-law density fluctuations,
the timescale for refractive scintillation ${\rm t_r}$ is given by 
\begin{equation}
{\rm t_r}\approx\frac{2\nu}{\Delta \nu_{\rm d}}\Delta t_{\rm d}
\label{eq:t_ref}
\end{equation}
\cite{ric90}. This parameter is given in column 8 of
Table~\ref{tb:para}. We use Equation~\ref{eq:cn2} to estimate mean
values of the fluctuation scaling parameter $C_{\rm n}^2$; derived values are
given in column 9 of Table~\ref{tb:para}.

Pulsar velocities can be estimated from the scintillation parameters
$\Delta t_{\rm d}$ and $\Delta\nu_{\rm d}$. For a thin scattering screen
placed midway between the pulsar and the observer, the transverse
velocity of the pattern is given by:
\begin{equation}
V_{\rm s} \approx 3.85 \times 10^4 \frac{\sqrt{\Delta\nu_{\rm d}\, D}}{\nu\,\Delta t_{\rm d}}
\label{eq:vs}
\end{equation}
where $V_{\rm s}$ is in km s$^{-1}$, $\Delta\nu_{\rm d}$ is in MHz,
the pulsar distance $D$ is in kpc, $\nu$ is in GHz, and $\Delta t_{\rm d}$
is in seconds \cite{grl94}. $V_{\rm s}$ is a combination of the
transverse velocity of the pulsar, the scattering medium and the
Earth's motion. The latter two components are usually comparatively
small, so $V_{\rm s}$ represents the transverse velocity of the
pulsar. From Equation~\ref{eq:vs} and the parallax distances given in
Table~\ref{tb:eph}, we derived transverse velocities for the five
pulsars.  The resulting values are plotted in Fig.~\ref{fg:0329_var}
to Fig.~\ref{fg:2021_var} and average values are given in column 10 of
Table~\ref{tb:para}.

\subsection{Flux densities}
\begin{figure}
\begin{center} 
\begin{tabular}{c}
\mbox{\psfig{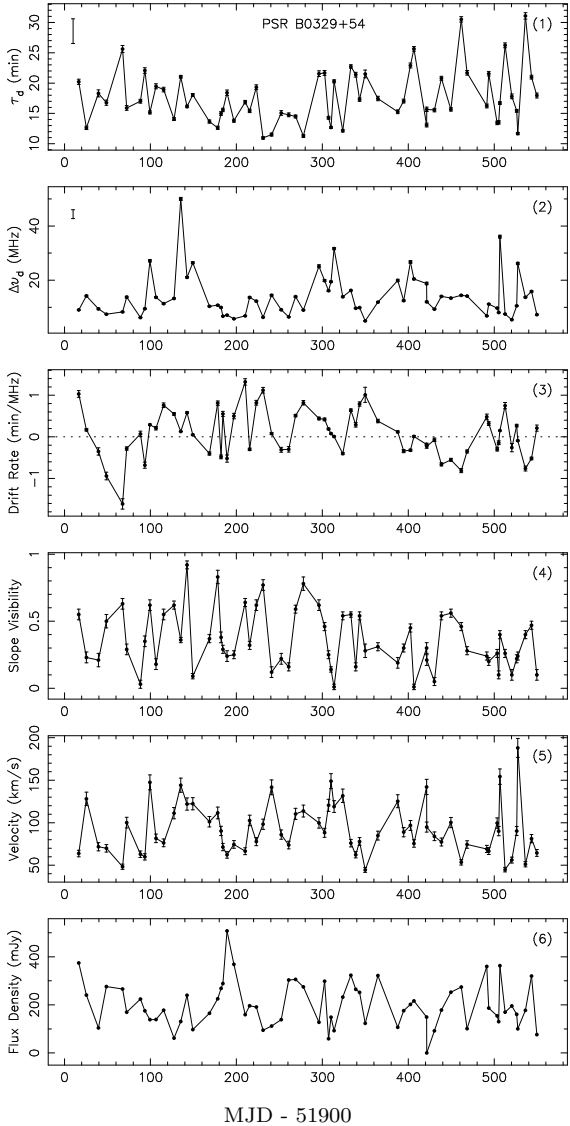}}\\
\vspace {-2mm}\\
MJD - 51900\\
\end{tabular}
\caption{Time variations of $\Delta t_{\rm d}$, $\Delta\nu_{\rm d}$, main
axial angle, slope visibility, transverse velocity and the flux density for
PSR B0329+54. Error bars ($\pm 1\sigma$) plotted on each point are
from the 2D-elliptical Gaussian fitting. The statistical uncertainty
related to the finite number of observed scintles is indicated by the
$\pm 1\sigma$ error bar in the mean value shown at the top-left corner
of the $\Delta t_{\rm d}$ and $\Delta\nu_d$ plots. }
\label{fg:0329_var}
\end{center}
\end{figure}

\begin{figure}
\begin{center} 
\begin{tabular}{c}
\mbox{\psfig{file=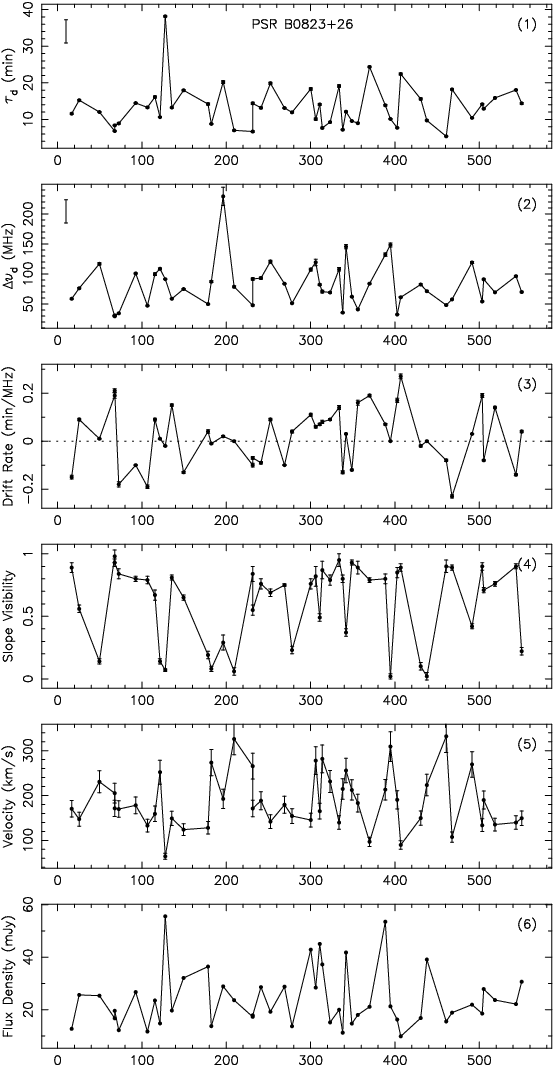,width=75mm}}\\
\vspace {-2mm}\\
MJD - 51900\\
\end{tabular}
\caption{As in Fig.~\ref{fg:0329_var}, time variations of $\Delta t_{\rm
d}$, $\Delta\nu_{\rm d}$, main axial angle, slope visibility,
transverse velocity and the flux density for PSR B0823+26.}
\label{fg:0823_var}
\end{center}
\end{figure}

\begin{figure}
\begin{center} 
\begin{tabular}{c}
\mbox{\psfig{file=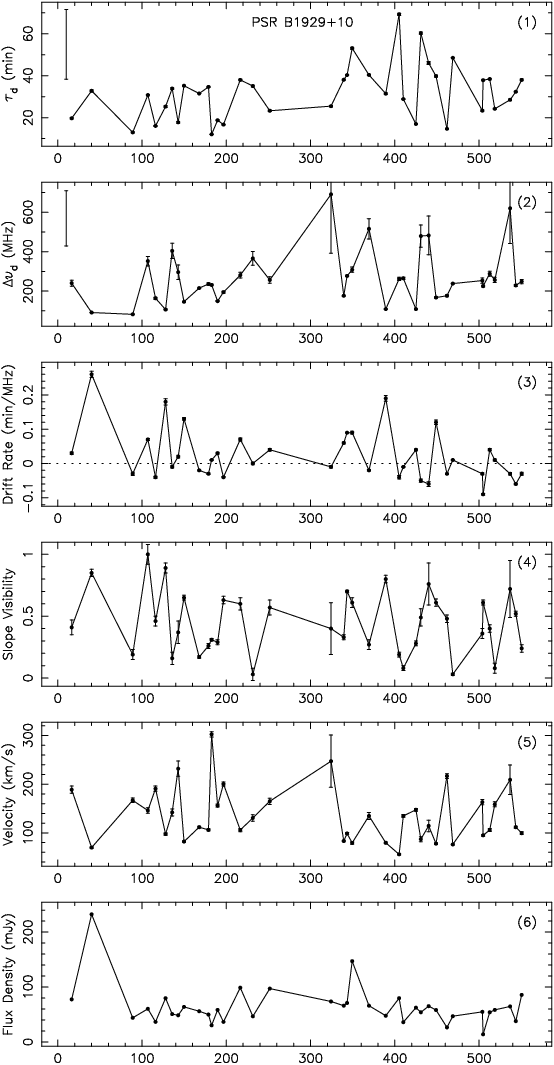,width=75mm}}\\
\vspace {-2mm}\\
MJD - 51900\\
\end{tabular}
\caption{As in Fig.~\ref{fg:0329_var}, time variations of $\Delta t_{\rm
d}$, $\Delta\nu_{\rm d}$, main axial angle, slope visibility,
transverse velocity and the flux density for PSR B1929+10.}
\label{fg:1929_var}
\end{center}
\end{figure}

\begin{figure}
\begin{center} 
\begin{tabular}{c}
\mbox{\psfig{file=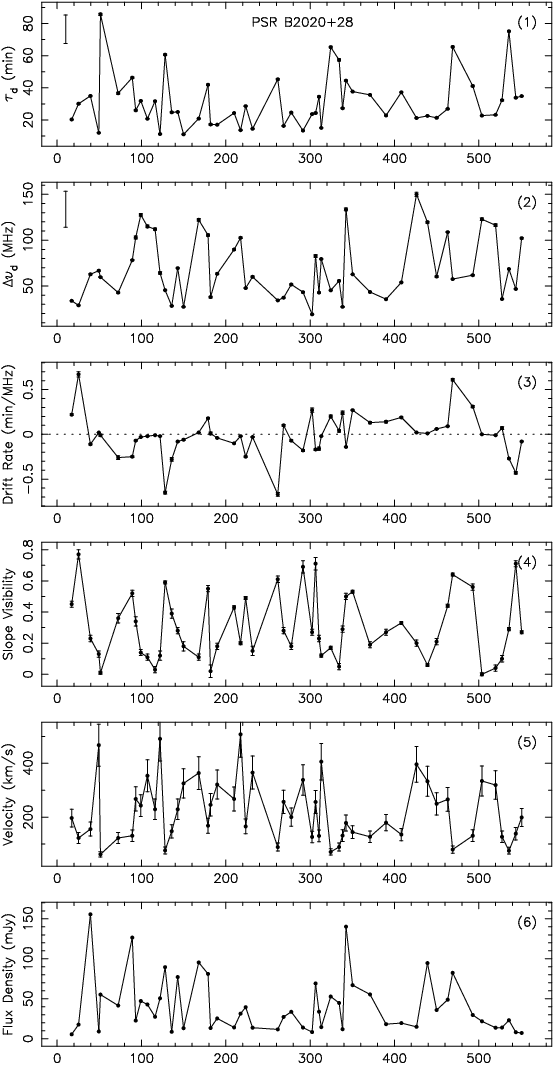,width=75mm}}\\
\vspace {-2mm}\\
MJD - 51900\\
\end{tabular}
\caption{As in Fig.~\ref{fg:0329_var}, time variations of $\Delta t_{\rm
d}$, $\Delta\nu_{\rm d}$, main axial angle, slope visibility,
transverse velocity and the flux density for PSR B2020+28.}
\label{fg:2020_var}
\end{center}
\end{figure}

\begin{figure}
\begin{center} 
\begin{tabular}{c}
\mbox{\psfig{file=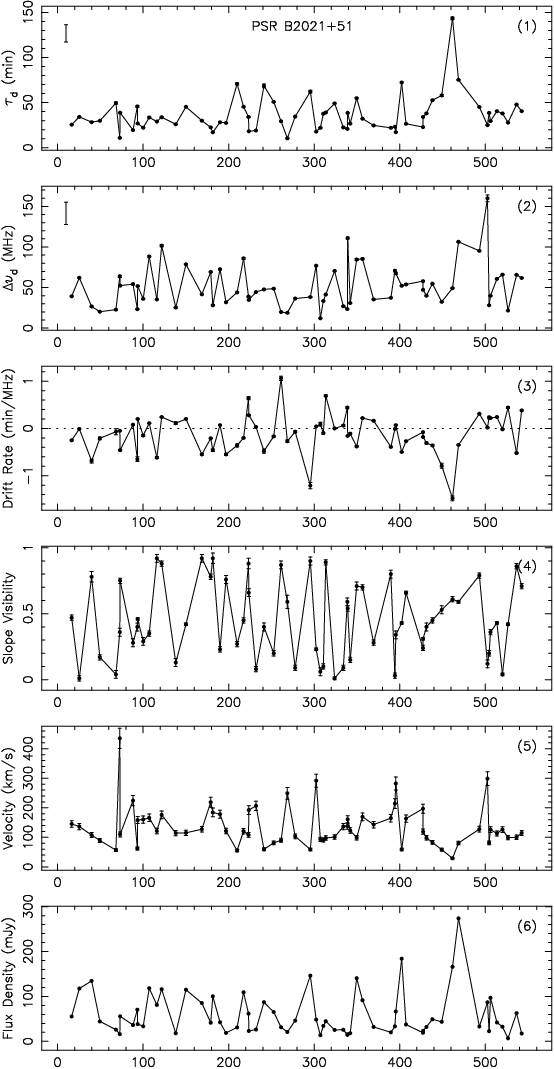,width=75mm}}\\
\vspace {-2mm}\\
MJD - 51900\\
\end{tabular}
\caption{As in Fig.~\ref{fg:0329_var}, time variations of $\Delta t_{\rm
d}$, $\Delta\nu_{\rm d}$, main axial angle, slope visibility,
transverse velocity and the flux density for PSR B2021+51.}
\label{fg:2021_var}
\end{center}
\end{figure}

Long-term observations have shown that pulsar flux density ($S$)
variations are dominated by refractive scintillation.  Distant pulsars
show little flux density variation, implying that intrinsic pulsar
radio luminosities are stable \cite{ks92}.  Observed flux densities
are plotted in Fig.~\ref{fg:0329_var} to Fig.~\ref{fg:2021_var} and
mean values are given in column 11 of Table~\ref{tb:para}.

\begin{table*}
\begin{minipage}{140mm}
\caption{Average scintillation parameters of the observed pulsars.}
\begin{tabular}{ccccccccccc}
\hline  & \vspace{-3mm} \\
PSR B   &$\Delta t_{\rm d}$ &$\Delta \nu_{\rm d}$  &
                          $\sigma_{\rm N}$
                                & \multicolumn{1}{c}{$dt/d\nu$}
                                            & \multicolumn{1}{c}{$|r|$} 
                                                      & $u$  & ${\rm t_r}$ 
                                                                       & $\log C_{\rm n}^2$ & $V_s$   & $S$\\
        & (min)    & (MHz)    &  &\multicolumn{1}{c}{(min/MHz)}  &    &      & (day)   &               &(km~s$^{-1}$) & (mJy)  \\
(1) & (2) & (3) & (4) & (5) & (6) & (7) & (8) & (9) & (10) & (11)\\
\hline  & \vspace{-3mm} \\ 
0329+54 & 16.9(5)  & 14(1)  &0.11 & ~0.07(7)   & 0.358(3)  & 15   & 2.5    &  $-3.0$        &~97(5)    &  ~199(11)\\ 
0823+26 & 13.6(8)  & 82(5)  &0.24 & ~0.02(2)   & 0.616(6)  & ~6   & 0.4    &  $-2.8$        &187(9)   &  ~~24(1)\\ 
1929+10 & 32(2)~~    & 268(24)&0.52 & ~0.02(1) & 0.442(7)  & ~3   & 0.3    &  $-3.1$        &136(9)   &  ~~65(5)\\ 
2020+28 & 31(2)~~    & 70(5)  &0.29 &--0.01(3) & 0.302(4)  & ~7   & 1.0    &  $-4.3$        &225(16)  &  ~~40(4)\\
2021+51 & 36(3)~~    & 52(3)  &0.26 &--0.11(5) & 0.452(4)  & ~8   & 1.5    &  $-4.0$        &138(8)   &  ~~59(6)\\ 
\hline  & \vspace{-3mm} \\
\end{tabular}
\label{tb:para}
\end{minipage}
\end{table*}

\section{Modelling the ACFs}\label{sec:modacf}
The observations of PSR B0329+54 gave us 64 independent estimates of
the frequency-time correlation function of intensity. Even though these
often show drifting features 
the average correlation function should still be estimated with good
confidence and we compared it with its theoretical shape (assuming
strong scattering) from a thin scattering medium with a Kolmogorov
spectrum.

\begin{figure}
\begin{center} 
\mbox{\psfig{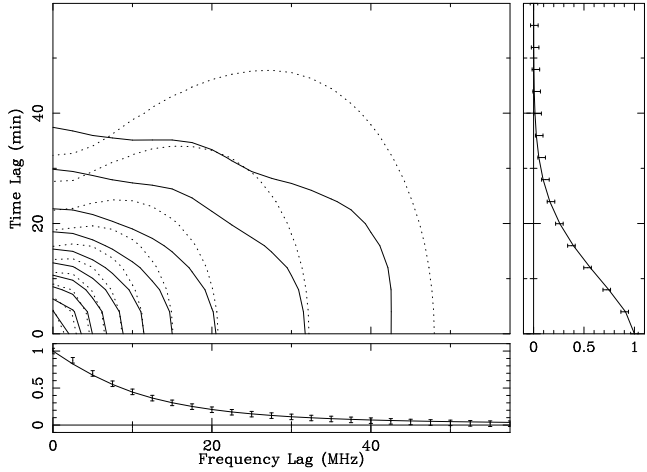}}
\caption{Auto correlation versus frequency and time lag for PSR
B0329+54.  Solid contours for the average auto correlation of 64
individual observations.  Dashed lines are contours for a Kolmogorov
model in the diffractive limit of scattering from a thin screen.
$\Delta \nu_{\rm d}$ and $\Delta t_{\rm d}$ were fitted to the cuts
along the axes as displayed in the two small bottom and right plots,
where the data are represented by error bars that are three times the
nominal standard error in the mean.  Notice how the dashed contours
agree with the solid contours on each axis, but bulge out away from
the axes.}
\label{fg:fit}
\end{center}
\end{figure}

The theoretical intensity correlation versus spatial and frequency
offsets, assuming a Kolmogorov spectrum and a thin screen, has been
derived by Lambert \& Rickett (1999; see their Figure 5)
\nocite{lr99}. In Fig.~\ref{fg:fit} we compare their model with the
mean of the observed auto-correlation functions, by assuming
that time offsets translate simply into spatial offsets. We estimated
$\Delta \nu_{\rm d}$ by fitting the theoretical shape of the cut along
the frequency axis and $\Delta t_{\rm d}$ from the cut along the time
axis.  The observed one-dimensional correlation functions are shown by
the error bars (three times the nominal standard error in the mean)
with a solid line for the model.

These plots show good general agreement in shape along both axes.
Particularly evident is the steeper slope near zero lag in the
frequency domain than in the time domain for both the model and the
observations.  The minimum chi-squared value is about twice the number
of data points in the fit, which maybe due to too low an
estimate of the error in the mean correlation function. We suspect
that the temporal variations in the two-dimensional correlations do
not follow normal statistics which causes the underestimate of the
error in its mean.

Whereas along the axes the model agrees with the theory, away from
the axes the theory contours bulge outward more than the observed
contours.  The outward bulging may be related to the arc phenomenon in
the secondary spectrum (see Section~\ref{sec:secsp}). We also note
that scattering extended along the line of sight causes a reduction in
the bulging effect, as can be seen in Figure 5 of Lambert \& Rickett
(1999).

\section{Secondary spectra}\label{sec:secsp}
\begin{figure}
\begin{center} 
\mbox{\psfig{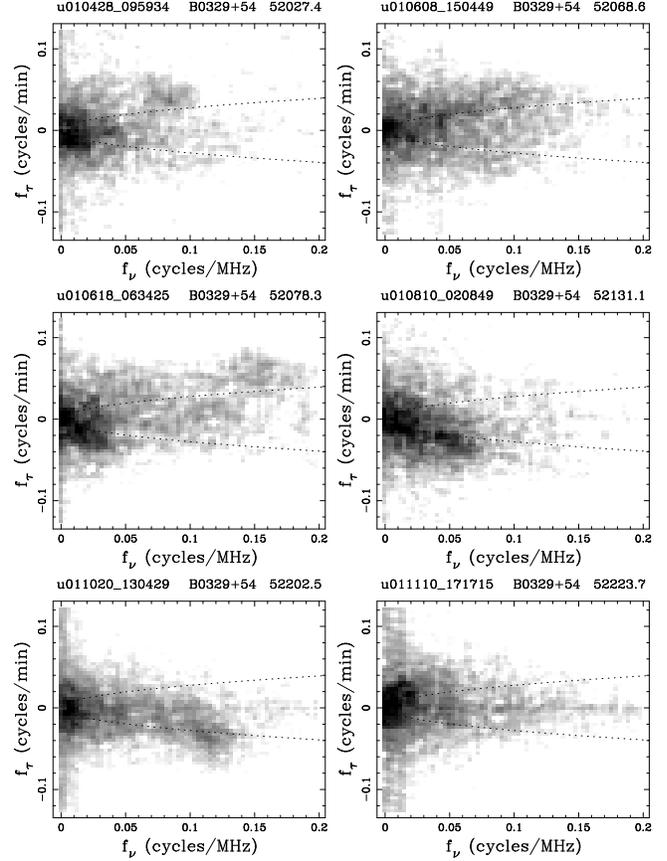}}
\caption{Secondary spectra for selected observations of PSR
B0329+54. The spectra have been Hanning weighted to reduce sidelobes
and improve the signal/noise ratio. The dotted lines indicate expected
arcs derived from Equation~8.}
\label{fg:2d}
\end{center}
\end{figure}

Secondary spectra, that is, two-dimensional Fourier transforms of
dynamic spectra, give additional insight into scintillation and
scattering in the interstellar medium. Fig.~\ref{fg:2d} shows
secondary spectra, for several observations of PSR B0329+54 where
significant fringing was observed. These spectra are much more
amorphous and centrally concentrated than those shown by
Hill et al. (2003)\nocite{hsb+03} for other pulsars at a similar
frequency. We note that their observations have narrower bandwidths
and shorter observation times but much higher time and frequency
resolution (10 -- 30 s and $\sim$0.1~MHz respectively) compared to
4~min and 2.5~MHz for our observations. Our observations therefore are
sensitive to larger-scale fringes and hence smaller values of
$\theta_r$ than those of Hill et al. (2003), i.e., we are observing
the central part of the scattering disk with high spatial resolution,
whereas the higher-resolution observations are senstive to large
refraction angles and very fine fringing structure. Although there are
some indications of rudimentary arc structure in Fig.~\ref{fg:2d}, the
model of interference between a strong central image and surrounding
discrete images is clearly less appropriate --- we are in fact
observing the structure of the central image.

Across our wide observed bandwidth ($\Delta\nu_{\rm obs} = 320$~MHz),
the frequency dependence of the slope ($dt/d\nu \propto \nu^{-3}$),
see Equation~\ref{eq:fringe}) corresponds to a 2:1 variation with
larger slopes (more vertical fringes) at lower frequencies. There is
some evidence for this in Fig.~\ref{fg:0329_spc}, for example, in the
narrow fringing visible in the spectrum for MJD 52202.5.  This slope
variation will smear the corresponding feature in the secondary
spectrum, contributing to the amorphous appearance of the secondary
spectra (Fig.~\ref{fg:2d}).

\section{Discussion}\label{sec:dis}
\subsection{Dynamic spectra}
For PSR B0329+54, the character of the dynamic spectra is very
different from that observed at lower frequencies. Sloping fringes of
different frequencies and slopes are very common in the 1540 MHz
observations, whereas at lower frequencies they are uncommon (e.g.,
Stinebring, Faison \& McKinnon, 1996). \nocite{sfm96}
The frequency dependence of decorrelation bandwidth is directly
observable within our receiver bandwidth in many dynamic
spectra. Taking the observation at MJD 52294.5 for PSR B0329+54 as an
example, the width of the scintle at 1400~MHz is about 13~MHz, whereas at
1640~MHz it is about 27~MHz, consistent with the
prediction from Equation~\ref{eq:dnu2}.

Despite PSRs B0329+54, B0823+26, B2020+28 and B2021+51 having very
similar dispersion measures, their scintillation properties are quite
different. As is well known, Galactic latitude appears to be important
with stronger scattering for PSR B0329+54. Distance seems less
important with very different DISS properties for PSR B0823+26 and PSR
B1929+10.

\subsection{Time variations}\label{sec:var}
Observed dynamic spectra vary greatly from observation to observation,
in all respects: decorrelation time scale, bandwidth, brightness and
the characteristics of sloping features. Fig.~\ref{fg:0329_var} to
Fig.~\ref{fg:2021_var} show that scintillation parameters have a wide
range of variation. For example, for PSR B0329+54
(Fig.~\ref{fg:0329_var}), the observed diffractive timescale and
decorrelation bandwidth vary over the ranges 12--30~min and 5--34~MHz
respectively. Similar variations are observed for the other four
pulsars. The interval between observations is typically several times
the refractive timescale ${\rm t_r}$ (Table~\ref{tb:para}) and so we
would expect uncorrelated variations. However, the expected modulation
due to refractive scintillation can only account for a small amount of
the observed variations. Modulation indices (ratio of rms deviation to
mean value) for the variations in $\Delta t_{\rm d}$, $\Delta\nu_{\rm
d}$ and flux density, $m_{\rm t}$, $m_{\rm b}$ and $m_{\rm r}$
respectively, are given in Table~\ref{tb:mod}. Also given in the Table
are predicted modulation indices for the case of a thin screen with a
Kolmogorov density spectrum ($\beta = 11/3$) and strong scattering
\cite{rnb86,bgr99}. In all cases, the observed modulation indices are
greater than the predicted values. With our relatively high observing
frequency and relatively nearby pulsars, the scattering
strength $u$ is only modestly greater than one, so the strong
scattering assumption is not really satisfied. Also, the form of the
variations suggests that large-scale fluctuations in the screen are
important, so the fluctuation spectrum is likely to be steeper than in
the Kolmogorov case.

\begin{table}
\caption{Modulation indices for $\Delta t_{\rm d}$, $\Delta\nu_{\rm
d}$ and flux density.}
\begin{tabular}{ccccccc}
\hline  & \vspace{-3mm} \\
PSR B   &$m_{\rm t}$ & $m_{\rm t\_th}$ & $m_{\rm b}$ & $m_{\rm b\_th}$ & $m_{\rm r}$ & $m_{\rm r\_th}$ \\
(1)     &  (2)   &  (3)    &  (4)   &  (5)   &  (6)  & (7) \\
\hline  & \vspace{-3mm} \\
0329+54 &  0.26  &  0.08   &  0.74  &  0.16  & 0.47  & 0.23 \\ 
0823+26 &  0.43  &  0.11   &  0.45  &  0.23  & 0.45  & 0.32 \\
1929+10 &  0.41  &  0.13   &  0.54  &  0.28  & 0.57  & 0.36 \\
2020+28 &  0.53  &  0.10   &  0.48  &  0.21  & 0.87  & 0.23 \\
2021+51 &  0.55  &  0.10   &  0.52  &  0.20  & 0.83  & 0.24 \\
\hline  & \vspace{-3mm} \\
\end{tabular}
\label{tb:mod}
\end{table}

Fig.~\ref{fg:0329_var} to Fig.~\ref{fg:2021_var} show that there are
significant correlated changes in decorrelation time-scale, bandwidth
and drift rate over intervals of several weeks and in some cases many
months, much larger than the predicted refractive timescale $t_{\rm
r}$. For example, in Fig.~\ref{fg:0329_var} for PSR B0329+54, there
are high values of decorrelation bandwidth at MJD 52035 and 52287, low
values of drift rate between MJD 52295 and 52370, an increase and then
decrease in slope visibility between MJD 52160 and 52215 and larger
than average flux densities between 52050 and 52110. These results
show that gradients in refractive index of scale much larger than that
of the diffractive scattering disk exist in the scattering screen.

In general, there is little correlation between variations in the
various parameters. However, there are exceptions. For example,
Fig.~\ref{fg:2021_var} shows a very significant correlation between
decorrelation time-scale and flux density. As remarked in
Section~\ref{sec:para}, the pulsar tends to be strong when the
scintillation timescale is long. This suggests that most of energy
from the pulsar is confined to the central region of the scattering
disk where the scattering angles are small, leading to broad scintles. Only
when we are in a null of these broad patterns do we see weaker
radiation from outer parts of the scattering disk where scattering
angles are larger and scintles are smaller.  

Due to the wide range variation of diffractive timescale and
decorrelation bandwidth, the derived velocity is also highly
variable. However, there is remarkably good agreement of the average
scintillation velocities (Table~\ref{tb:para}) with the velocities
derived from proper motion measurements (Table~\ref{tb:eph}). We
attribute this to our long-term observations which averaged over the
refractive scintillation modulations \cite{brg99}. This agreement also
indicates that locating a thin screen approximately half-way between
the pulsar and observer is a good assumption \cite{grl94,nnd+01}.

\subsection{Power spectrum of electron density turbulence}
The frequency dependence of the DISS parameters can be used to
estimate the spectral slope of interstellar turbulence.
Table~\ref{tb:scale} shows the results from earlier low-frequency
observations, with the data in columns 2 and 3 derived from observations using
the Ooty Radio Telescope at 327~MHz by Bhat, Rao \& Gupta
(1999), \nocite{brg99} columns 4 and 5 are from 408~MHz
observations by Gupta, Rickett \& Lyne (1994)\nocite{grl94} and columns 6
and 7 are from 610~MHz observations by Stinebring, Faison \& McKinnon
(1996).\nocite{sfm96} For convenience, columns 8 and 9 are from our
observations, repeated from Table~\ref{tb:para}. For Kolmogorov
turbulence, the frequency dependences of scintillation parameters are
given by: $\Delta t_{\rm d}\propto\nu^{1.2}$, $\Delta\nu_{\rm
d}\propto\nu^{4.4}$ (see Equations~\ref{eq:td2} and \ref{eq:dnu2}).
Scintillation timescales are in reasonable accord with the predicted
frequency dependence, but the observed values of the decorrelation
bandwidth are systematically lower at high frequencies. Fitting for
the observed time scale and decorrelation bandwidth we obtained the
power index, $\alpha_t$ and $\alpha_\nu$ and the corresponding density
spectral indices $\beta_t$ and $\beta_{\nu}$, using a model for a
density fluctuation spectrum with $\beta<4$ (Equations~\ref{eq:td1} \&
\ref{eq:dnu1}). These values, shown in the last four columns in
Table~\ref{tb:scale} (except in one case) show a flatter frequency
dependence for $\Delta t_{\rm d}$ and $\Delta \nu_{\rm d}$ than
predicted, implying a density fluctuation index much larger than the
Kolmogorov value and in fact greater than the critical value of
four. We therefore applied a model for a steeper density fluctuation
spectrum, $\beta>4$ (Equations~\ref{eq:td11} \&
\ref{eq:dnu11}). However, the derived values contradict the assumption
giving $\beta_t<4$ and $\beta_{\nu}<4$.

\begin{table*}
\begin{minipage}{140mm}
\caption{Fitting for frequency dependences of $\Delta t_{\rm d}$ and
$\Delta\nu_{\rm d}$ and the corresponding density spectra.}
\begin{tabular}{ccccccccccccc}
\hline  & \vspace{-3mm} \\
Obs. Freq& \multicolumn{2}{c}{327~MHz\footnote{Bhat, Rao \& Gupta 1999}}
         & \multicolumn{2}{c}{408~MHz\footnote{Gupta, Rickett \& Lyne 1994}}
         & \multicolumn{2}{c}{610~MHz\footnote{Stinebring, Faison \& McKinnon 1996}}
         & \multicolumn{2}{c}{1540~MHz\footnote{Our Results}} & & & &\\
--------------  & ~~$\Delta t_{\rm d}$  & ~$\Delta\nu_{\rm d}$
        & $\Delta t_{\rm d}$  & $\Delta\nu_{\rm d}$ 
        & $\Delta t_{\rm d}$  & $\Delta\nu_{\rm d}$ 
        & $\Delta t_{\rm d}$  & $\Delta\nu_{\rm d}$ & $\alpha_t$  & $\beta_t$  & $\alpha_{\nu}$  & $\beta_{\nu}$\\
PSR B   & (min)           & (MHz)   
        & (min)           & (MHz)   
        & (min)           & (MHz) 
        & (min)           & (MHz)  & & & & \\
    (1) &   (2)&  (3)     &  (4)   &  (5)  & (6)    &   (7)  & (8)     & (9)   & (10)& (11) & (12) & (13)\\
\hline  & \vspace{-3mm} \\
0329+54 & 5.1  & 0.165    &  3.2   & 0.046 &  5.9   & 0.349  & 16.9    & 14    & 0.93 & 4.2 & 3.3 & 5.0\\
0823+26 & 4.2  & 0.293    &  1.7   & 0.269 &        &        & 13.6    & 82    & 1.00 & 4.0 & 3.9 & 4.1\\
1929+10 & 5.8  & 1.293    &        &       &        &        & 32      & 268   & 1.10 & 3.8 & 3.4 & 4.8 \\
2020+28 & 4.7  & 0.270    &        &       &        &        & 31      & 70    & 1.22 & 3.6 & 3.6 & 4.5\\
2021+51 &      &          &        &       &        &        & 36      & 52    &     & & & \\
\hline  & \vspace{-3mm} \\
\end{tabular}
\label{tb:scale}
\end{minipage}
\end{table*}

The flat frequency dependence of $\Delta t_{\rm d}$ and $\Delta
\nu_{\rm d}$ is not predicted by either theory. Similar low values of
$\alpha_{\nu}$ were also observed earlier by Johnston, Nicastro \&
Koribalski (1998) and L\"ohmer et al. (2001). The former work consists
of multi-frequency DISS observations for large sample of pulsars and
in the latter paper the index was derived from scatter broadening
observations. This result may imply a relatively small outer scale for
the scattering fluctuations or may be a consequence of weaker
scattering at the higher frequencies.

\nocite{jnk98} \nocite{lkm+01}

By comparing the diffractive and refractive scattering angles,
$\theta_{\rm d}$ and $\theta_{\rm r}$ respectively, we can
estimate the spectral index $\beta$ of the density fluctuations
(Equation~\ref{eq:kolmogorov}). In terms of observable quantities,
they are given by

\begin{equation}
\theta_{\rm d}=\left(\frac{c}{\pi \,D \,\Delta \nu_{\rm d}}\right)^{0.5},
\label{eq:theta_d}
\end{equation} 
and Equation~\ref{eq:fringe} rearranged
\begin{equation}
\theta_{\rm r}=\left(\frac{V_{\rm s}\,\nu}{D}\right)\left(\frac{dt}{d\nu}\right),
\label{eq:theta_r}
\end{equation} 
where $V_{\rm s}$ is given in Table~\ref{tb:para} and $dt/d\nu$ is the
drift slope of the dynamic spectrum from Equation~\ref{eq:slope}. The
mean values of $\langle\theta_{\rm d}\rangle$ and $\langle\theta_{\rm
r}\rangle$ are given in columns 2 and 3 of Table~\ref{tb:theta}. As
expected for random fluctuations in the scattering medium,
$\langle\theta_{\rm r}\rangle$ is essentially zero within the
uncertainties. However, the scale of the refractive scattering can be
estimated from the rms fluctuation of $\theta_{\rm r}$, denoted as
$\sigma_{\theta_{\rm r}}$ and given in column 4 of
Table~\ref{tb:theta}. Following Bhat, Gupta \& Rao (1999) we scale
this quantity by a factor of $\sqrt{2}$ to allow for mis-alignment of
the refractive gradient and the pattern velocity.

We estimate the spectral index of the electron density fluctuations using
\begin{equation}
\beta=4+\left[\frac{\log(\sigma_{\theta_{\rm r}}/\langle\theta_{\rm d}\rangle)}{\log u}\right],
\label{eq:beta}
\end{equation}
where $u$ is the measurement of scattering strength defined in
Equation~\ref{eq:u} \cite{bgr99}. This formula indicates that
$\sigma_{\theta_{\rm r}}<\langle\theta_{\rm d}\rangle$
corresponds to a flatter spectrum $\beta<4$ but $\sigma_{\theta_{\rm
r}}>\langle\theta_{\rm d}\rangle$ corresponds to a steeper spectrum
$\beta>4$. In our case as $\sigma_{\theta_{\rm r}} \ll
\langle\theta_{\rm d}\rangle$, and the measured $\beta$ are in the range
3.40 -- 3.75, on average somewhat less than the Kolmogorov spectrum index of
3.67, but in most cases consistent with it. 

Taken at face value, these results are inconsistent with those based
on the frequency dependence of scintillation parameters
(Table~\ref{tb:scale}). One possible explanation is that the
fluctuations at the diffractive and refractive scales are independent,
not related by the cascade of turbulent energy and hence not described
by the Kolmogorov spectrum \cite{ars95}.  We also note that our range
of derived drift slopes is limited by the observing parameters,
especially time and frequency resolution. Higher resolution
observations (e.g. Hill et al. 2003) detect a wider range of slopes
and hence imply larger refractive angles and a steeper spectrum for
the density fluctuations.

\begin{table}
\begin{minipage}{65mm}
\caption{Estimates of diffractive and refractive scattering angles,
and the spectral index of electron density fluctuation.}
\begin{tabular}{ccrcc}
\hline  & \vspace{-3mm} \\
PSR B &$\langle\theta_{\rm d}\rangle$&$\langle\theta_{\rm r}\rangle$~~~~~&$\sigma_{\theta_{\rm r}}$&$\beta$ \\
        &     (mas)       &      (mas)~~     &        (mas)    &         \\
(1)     &    (2)          &      (3)~~~    &       (4)       &   (5)   \\
\hline  & \vspace{-3mm} \\
0329+54 &    0.106(3)     &     0.005(3)   &       0.036     &    3.59 \\
0823+26 &    0.070(2)     &     0.004(5)   &       0.044     &    3.75 \\
1929+10 &    0.043(2)     &     0.003(2)   &       0.019     &    3.33 \\
2020+28 &    0.029(1)     &   $-$0.0003(9) &       0.009     &    3.40 \\
2021+54 &    0.039(1)     &   $-$0.002(2)  &       0.017     &    3.61 \\
\hline  & \vspace{-3mm} \\
\end{tabular}
\label{tb:theta}
\end{minipage}
\end{table}

\section{Summary}\label{sec:sum}
In this paper we present dynamic spectra for PSRs B0329+54, B0823+26,
B1929+10, B2020+28 and B2021+51 over a 320~MHz centred at
1540~MHz. The time and frequency bandwidth resolutions are 4~min and
2.5 MHz respectively. The dynamic spectra show different features
compared to earlier lower-frequency observations. For example, PSR
B0329+54 presents frequent drifting patterns and fringes which were
not seen at lower frequencies. Our individual observations are
generally long enough to cover at least a few scintles, but in some
cases, especially for PSR B1929+10, the scintles are wide compared to
our bandwidth. In these cases, the value of $\Delta\nu_{\rm d}$ is
based on just the peak of the ACF and is rather uncertain. Although
the observations show a wide range of variations in diffractive
scintillation parameters, the average scintillation velocities are in
good agreement with the proper motion velocities.


The frequency and time correlations are successfully described by
theory for a thin Kolmogorov scattering screen.  However, there is a
consistent disagreement for points off the axes, which suggests
turbulence extended along the line of sight.  There is also much
greater variability than expected in the shapes of the daily
auto-correlations, which suggests a stronger than expected influence
of scales larger than the scattering disc. We note here that similar
conclusion has been drawn by Lambert \& Rickett (2000), Spangler
(2001) and Shishov et al. (2003). These authors also pointed to
more complex models than a Kolmogorov thin screen. We do not have
theoretical predictions for the full frequency-time correlation for
the implied steeper spectrum models, which thus becomes an area for
future work.

\nocite{lr00,spa01,sss+03}

With our wide observed bandwith, long observation times and lower time
and frequency resolution, our observations are sensitive to
larger-scale fringes and hence smaller refractive angles,
corresponding to the central part of the scattering disk. The
secondary spectra we obtained for PSR B0329+54 are centrally
concentrated and rather amorphous. Some of the smearing results from a
$\nu^{-3}$ frequency dependence of fringe slope across our wide
observed band.

Modulations in time scale, decorrelation bandwidth and flux density
are greater than predictions from a thin screen model with a
Kolmogorov spectrum of density fluctuations. The long-term variations
in DISS parameters suggests larger-scale density fluctuations exist,
corresponding to a steeper fluctuation spectrum.  Comparison of the
observed DISS timescales with values extrapolated from lower frequency
measurements based on the Kolmogorov spectrum show good agreement.
However, large discrepancies were found in extrapolated decorrelation
bandwidths, with the observed values substantially smaller than the
predicted values. This gives a flatter frequency dependence of
decorrelation bandwidth, corresponding to a steeper density
fluctuation spectrum.

In contrast, estimation of the power-law density fluctuation index
$\beta$ from diffractive and refractive scattering angles gives a
value in the range 3.40 -- 3.75, on average smaller but consistent
with a Kolmogorov spectrum. The range of derived drift slopes is
limited by the observing parameters, and observations with higher
resolution will detect larger refractive angles and hence imply a
steeper density spectrum. It is also possible that the large-scale
fluctuations responsible for refractive effects are not directly
related to the smaller-scale fluctuations responsible for diffractive
scintillation.

\section*{ACKNOWLEDGMENTS}
We thank the engineers of Urumqi Observatory who help with maintaining
the system, we also thank those who helped with the observations. NW
thanks the support from NNSFC under the project of 10173020. We thank
Mark Walker for his helpful suggestions and discussions.
 
\bibliographystyle{mn}
\bibliography{journals,psrrefs,modrefs,crossrefs,nwrefs}

\end{document}